\documentclass[prb,twocolumn,superscriptaddress,amsmath,amssymb,aps]{revtex4-2}

\usepackage{color}
\usepackage{graphicx}
\usepackage{dcolumn}
\usepackage{bm}
\usepackage{hyperref}

\newcommand{\ua}{\uparrow} 
\newcommand{\da}{\downarrow}

\newcommand{\bk}{\mathbf k}

\newcommand{\bq}{\mathbf q}
\newcommand{\br}{\mathbf r}

\newcommand{\be}{\boldsymbol e}

\newcommand\mk{\boldsymbol{\mathrm{k}}}
\newcommand{\bDelta}{\boldsymbol{\Delta}}
\newcommand{\rvec}[1]{\left| #1 \right\rangle}
\newcommand{\lvec}[1]{\left\langle #1 \right|}
\newcommand{\vecDot}[2]{\left \langle #1 \middle| #2 \right \rangle}
\newcommand{\abs}[1]{\left| #1 \right|}
\newcommand{\ti}[1]{\scriptscriptstyle{#1}}

\usepackage[normalem]{ulem}

\begin{document}

\title{Pair density waves via interband pairing --- a perspective from superfluid weight}

\author{Yi-Jian Hu}
\affiliation{Southern University of Science and Technology, Shenzhen 518055, Guangdong, China}
\affiliation{School of Physical Sciences, Great Bay University, Dongguan 523000, China}
\affiliation{International Quantum Academy, Shenzhen 518048, China}

\author{Yi-Fan Jiang}
\affiliation{State Key Laboratory of Quantum Functional Materials, School of Physical Science and Technology, ShanghaiTech University, Shanghai 201210, China}

\author{Wen Huang}%
\email{huangwen@gbu.edu.cn}
\affiliation{School of Physical Sciences, Great Bay University, Dongguan 523000, China}
\affiliation{International Quantum Academy, Shenzhen 518048, China}

\date{\today}

\begin{abstract}
Pair density waves (PDW) are superconducting states that exhibit a spatially modulated pairing order parameter, corresponding to Cooper pairs carrying finite center-of-mass momentum. While some recent studies have emphasized the role of quantum geometry in stabilizing PDW order in certain multiband systems, we here present a systematic investigation of an alternative, geometry-independent mechanism: PDW order driven purely by interband Cooper pairing. Using the superfluid weight (SW) as a diagnostic, we study two canonical models. The first is the classic Fulde-Ferrell model of an $s$-wave superconductor subject to a Zeeman splitting field, which realizes pure interband pairing between two rigidly shifted bands; the second is a spinless honeycomb-lattice model with nearest-neighbor $f$-wave pairing, a modified version of which was recently found to exhibit PDW instability in density-matrix renormalization-group simulations. In both models the SW receives no quantum geometric contribution at the mean-field level, yet it can be greatly suppressed or even turn negative, signaling a tendency toward pairing with spatial phase modulations. For the first model, we establish that the SW is increasingly suppressed as the Zeeman field grows, drastically turning negative upon the emergence of extensive gapless excitations, i.e., the Bogoliubov Fermi surfaces. In conjunction, our calculations also confirm three representative free-energy profiles as a function of the Cooper-pair modulation wavevector, characterized by whether uniform pairing corresponds to free energy global minimum, local maximum, or local minimum. For the honeycomb $f$-wave model, we demonstrate, both analytically and numerically, vanishing quantum-geometric SW and an invariably negative total SW. Our results thus establish the SW as a qualitative indicator of the tendency toward PDW order, and provide a microscopic rationale for the interband-pairing-driven PDW instability.
\end{abstract}

\maketitle


\section{Introduction}
The pair density wave (PDW) is a distinct class of superconducting order in which the superconducting order parameter undergoes periodic spatial modulation, corresponding in momentum space to Cooper pairs with finite center-of-mass momentum. While a foundational theory of such finite-momentum pairing was established by Fulde, Ferrell, Larkin, and Ovchinnikov (FFLO) in the context of Zeeman-field-induced Fermi surface splitting in conventional $s$-wave superconductors~\cite{fulde1964superconductivity,larkin1965nonuniform}, recent experimental advances have revealed that PDW order can emerge intrinsically at zero magnetic field without violating time-reversal symmetry. This intrinsic PDW is notably implicated in the pseudogap physics of underdoped cuprates and has been experimentally identified across a diverse range of unconventional superconductors, including iron-based systems~\cite{liuPairDensityWave2023,zhang2024visualizing}, UTe$_2$~\cite{gu2023detection},  $A$V$_3$Sb$_5$~\cite{chen2021roton,deng2024chiral}, and transition-metal dichalcogenides~\cite{Xiaolong2021Discovery_Science, LiXuan2025Unidirectional}. Despite these experimental breakthroughs, a comprehensive microscopic understanding of how PDW order arises from strong electron correlations remains an outstanding theoretical challenge\cite{Berg2010,Jaefari2012,SKJian-2015PRL,ZYHan-2020PRL,Shaffer2023Triplet,Shaffer2023Weak,Wu2023PairPRB,Castro2023Emergence,liu2024enhanced,ticea2024pair,Chen2023Singlet,Tsvelik2023,Wu2023Sublattice,Han2022Pair,yue2024pseudogap,Donna:23dwave,Patrick2014Amperean,Wu2023Pair,Haokai2023,Setty2023Exact,Jiang2023Pair,JiangAndTom2023Pair-Frontiers,Jin2022PRL,Banerjee2022Charge,Song2022Doping,GuoHM2024,Kivelson2020NPJQuantumM, Jiucai2025Pair, Yifan2024Pair, Yahui2022Pair,wang2026anomalous,zhang2025spin-triplet}. 

\begin{figure}
    \centering
    \includegraphics[width=\columnwidth]{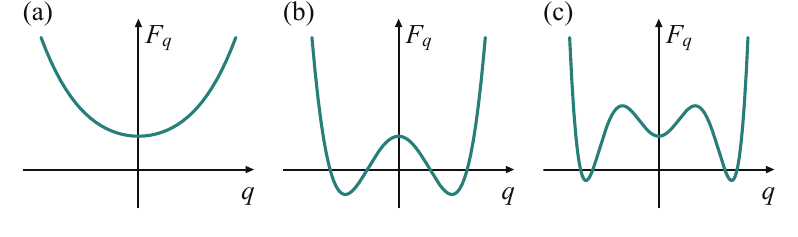}
    \caption{Schematic of the representative free-energy profiles as a function of Cooper pairing momentum $q$, illustrating different possible free energy profiles characterizing three different scenarios of the SW and ground state. (a) $D>0$, uniform state is global free-energy minimum. (b) $D<0$, PDW state is favored, and uniform pairing corresponds to local energy maximum. (c) $D>0$, PDW state is favored, and uniform pairing is a local energy minimum.}
    \label{fig:three_modes}
\end{figure}
The superfluid weight (SW) serves as a measure of the energy cost of a phase modulation in a superconductor. Specifically, the SW of a uniform superconducting order can be defined as the second derivative of the free energy with respect to the modulation wavevector $\bq$ of the pairing phase $e^{i\bq\cdot\br}$:
\begin{equation}\label{eq:Fq2ndderivative}
    D_{\mu\nu}=4\left.\frac{\partial^2 F_{\bq}}{\partial q_{\mu}\partial q_{\nu}}\right|_{\bq=0},
\end{equation}
where $\mu,\nu=x,y$ denotes the direction. In isotropic models, the tensor $D_{\mu\nu}$ is diagonal and can be treated as a scalar $D$. When the SW is positive, phase modulation raises the energy of the system, indicating that the uniform superconducting state is energetically favorable (Fig.~\ref{fig:three_modes} (a)). This is the case for most conventional superconductors under zero Zeeman splitting field. In the second scenario, when $D<0$, the system may favor a state exhibiting phase modulations, leading to the condensation of a PDW order (Fig.~\ref{fig:three_modes} (b)). However, the above two scenarios are not the only possibilities. In a third scenario, the SW can be positive and $\bq=0$ is a local free-energy minimum, yet the global minimum occurs at some finite $\bq$, as illustrated in Fig.~\ref{fig:three_modes} (c). 

Recent advances have found that the SW in multiband systems generally receive a contribution closely related to the quantum geometric character of the paired electrons~\cite{peottaSuperfluidityTopologicallyNontrivial2015,liangBandGeometryBerry2017,Hu2025Quantumgeometric,tormaSuperconductivitySuperfluidityQuantum2022,julkuGeometricOriginSuperfluidity2016,iskin2018exposing,iskin2019superfluid,iskin2019origin,huGeometricConventionalContribution2019,julkuSuperfluidWeightBerezinskiiKosterlitzThouless2020,xieTopologyBoundedSuperfluidWeight2020,wang2020quantum,herzog-arbeitmanSuperfluidWeightBounds2022,huhtinenRevisitingFlatBand2022,kitamura2022superconductivity,jiang2024geometric,jiang2025superfluid,iskin2024cooperpairing,peottaQuantumGeometrySuperfluidity2023,han2024quantum,tian2023diracflatband,tanaka2025superfluidstiffness}. This contribution may become negative under certain circumstances, thus providing a potential intrinsic mechanism to develop PDW order ~\cite{kitamuraQuantumGeometricEffect2022,chenPairDensityWave2023,jiangPairDensityWaves2023,ticea2024pair,wang2024quantum,lamponen2025superconductivity,sun2025flat}. Separately, in a previous work ~\cite{wang2025negativeroutepairdensity}, some of the authors identified an alternative mechanism that can also generate negative SW in multiband systems without relying on the quantum geometric effects, but it involves interband Cooper pairing between bands that disperse in opposite directions.   

In this work, we present a more thorough investigation of the interband-pairing-driven PDW order free of quantum-geometric effects. Based on studies of two canonical models, our aim is to establish a qualitative--sometimes even quantitative--connection between the mean-field SW and the tendency to develop PDW order. The first model we shall study is a conventional $s$-wave superconductor subject to an external Zeeman splitting field, which is in fact a classic model that exhibits only interband pairing as well as vanishing geometric effect. Here, the Zeeman-split up and down spin species can be viewed as two rigidly shifted, fully decoupled bands. While the FFLO states have been studied extensively in the past, they have not been carefully analyzed through the lens of the SW, and it is not clear whether the free-energy profile belongs to any of the latter two scenarios mentioned above. In this work, we evaluate the SW of this model and investigate the condition for the formation of Fulde-Ferrell (FF) type PDW order. The second model is a spinless honeycomb lattice model with only nearest-neighbor hopping and nearest-neighbor $f$-wave pairing, similar to a model recently found to support PDW order by density-matrix-renormalization-group (DMRG) calculations~\cite{Yifan2024Pair}. Within the mean-field framework, this model turns out to be one that has only interband pairing. We shall find that its mean-field SW has no geometric contribution and is negative, hence establishing a tendency toward PDW order. 

The remainder of this paper is organized as follows. Section \ref{sec:SW} presents the general formalism for evaluating the SW of uniform superconducting pairings. In Sec. \ref{sec:FFPDW}, we study the evolution of the SW of the conventional superconductor under the influence of a Zeeman splitting field and investigate its relation with the formation of the conventional FF-type PDW phase. In Sec. \ref{sec:Honeycomb}, we focus on the aforementioned honeycomb $f$-wave model and demonstrate that negative SW is a prelude to the formation of PDW order. Section \ref{sec:summary} provides a summary of this study. In the Appendix, we provide a detailed derivation of the SW in superconductors with pure interband pairing, along with an application to the honeycomb $f$-wave model. 

\section{Superfluid weight}
\label{sec:SW}
In general, the SW can be derived from the standard linear response theory~\cite{liangBandGeometryBerry2017},
\begin{equation}\label{eq:SW_formula}
    D_{\mu \nu}=\frac{T}{\Omega} \sum_{\omega_n, \mk} \operatorname{Tr}\left\{\tilde{m}_{\mu \nu\bk}^{-1} \mathcal{G}_{\bk}+\mathcal{G}_{\bk}\tilde{V}_{\mu\bk} \mathcal{G}_{\bk}\tilde{V}_{\nu\bk}\right\},
\end{equation}
where $T$ represents temperature, $\omega_n =(2n+1)\pi T$ denotes fermionic Matsubara frequency, $\Omega$ is the system size, and $\mathcal{G} \equiv \mathcal{G}(\mathrm{i}\omega_n,\mk)=(i\omega_n - \mathcal{H}_{\mk}^\text{BdG})^{-1}$ is the Gor'kov Green's function associated with the Bogoliubov-de Gennes (BdG) Hamiltonian. The inverse mass tensor and the velocity operator are defined as follows,
\begin{align}
\tilde{m}_{\mu\nu\bk}^{-1}&=\left.\partial_{\mu}\partial_{\nu}\mathcal{H}_{\mk}^{\text{BdG}}\right|_{\Delta\to 0},\\
\tilde{V}_{\mu\bk}&=\left.\tau_z \partial_{\mu}\mathcal{H}_{\mk}^{\mathrm{BdG}}\right|_{\Delta\to 0},\label{eq:Vorbital}
\end{align}
where $\tau_{i}(i=x,y,z)$ represents the Pauli matrices in the Nambu particle-hole space. Here and after, we use the shorthand notation $\partial_{\mu} \equiv \frac{\partial}{\partial k_{\mu}}$ and $\bar{\mk}\equiv-\mk$. In a general multiband system, the velocity matrix consists of both the intraband velocity $V^{ii}_{\mu}=\partial_\mu\varepsilon_i$ and the interband velocity $V^{ij}_{\mu}=(\varepsilon_i-\varepsilon_j)\vecDot{\partial_{\mu}i}{j}$, where $\varepsilon_i$ and $\rvec{i}$ are the energy dispersion and Bloch wavefunction of band-$i$, respectively. Here, the explicit dependence on $\bk$ is suppressed for brevity. The intraband velocity is just the conventional group velocity of the bands, whereas the interband velocity is closely tied to the quantum geometric character of the Bloch states~\cite{niuqianRMP,liu2024quantum,yu2025quantumgeometry,gao2026quantum}. Any contribution to the SW involving interband velocity is considered geometric in nature.

In the simple example of a single-band $s$-wave superconductor, a formal derivation according to Eq.~\eqref{eq:SW_formula} leads to~\cite{Hu2025Quantumgeometric}
\begin{equation}\label{eq:convSW}
    D_{\mu\nu} = -\frac{1}{\Omega}\sum_{\mk}\left( \frac{f(-E_{\bk})}{E_{\bk}}+2f'(-E_{\bk}) \right)\frac{V_{\mu\mk} (V_{\nu\bar{\mk}})^\ast\Delta\Delta^\ast }{(E_{\mk})^2},
\end{equation}
where $E_{\mk}=\sqrt{\varepsilon_{\mk}^2+|\Delta|^2}$ is the Bogoliubov quasiparticle dispersion, and $f'(E_{\bk})$ is the temperature-derivative of Fermi distribution function. In systems with time-reversal symmetry, $V_{\mu\mk}=-(V_{\mu\bar{\mk}})^*$ holds, thus the first term in the parentheses is positive definite. The second term constitutes a negative contribution arising from thermal excitations of quasiparticles. At $T=0$, $f'(E_{\bk})\approx -\delta(E_{\bk})$, hence this term vanishes in a fully gapped superconducting state, whereas it produces a negative contribution in a gapless state. In the special case of a state exhibiting extensive nodal excitations---such as Bogoliubov Fermi surfaces---the net SW may be severely suppressed, potentially destabilizing the uniform superconducting phase.  

On the other hand, for a simple two-band model with vanishing interband velocity and with pure $\bk$-independent interband pairing, i.e., $\Delta_{ij} c_{i\bar{\mk}\ua}c_{j\mk\da}$, the expression of the SW involves the following terms in the integrand (see Appendix~\ref{subsec:conv} for more details)
\begin{equation}\label{eq:sw_interband}
\begin{aligned}
    &\left[-\frac{f(E_{\alpha,\bk}^+)-f(E_{\alpha,\bk}^-)}{g_{\alpha,\bk}} + f'(E_{\alpha,\bk}^+)+f'(E_{\alpha,\bk}^-)\right]\\
    &\qquad\qquad\qquad\qquad\times V^{ii}_{\mu\mk}V^{jj}_{\nu\mk}\Delta_{ij}^\ast \Delta_{ij},
    \end{aligned}
\end{equation}
where $V^{ii}_{\mu\bk}$ denotes the $\mu$-th component of band-$i$'s group velocity, $g_{\alpha,\bk}=\sqrt{\frac{1}{4}(\varepsilon_{1\bk}+\varepsilon_{2\bk})^{2}+\abs{\Delta_{12,\bk}}^2}$ and $\varepsilon_{i\bk}$ is the dispersion of band-$i$. Two of the quasiparticle bands disperse according to
\begin{equation}
    E^{\pm}_{\alpha,\bk}=\frac{\varepsilon_{1\bk}-\varepsilon_{2\bk}}{2}\pm g_{\alpha,\bk},
\end{equation}
while the other two quasiparticle bands are opposite to $E^{\pm}_{\alpha,\bk}$ if $\Delta_{12}=\abs{\Delta_{21}}$. Interestingly, the sign of $V^{ii}_{\mu\mk}V^{jj}_{\nu\mk}$ depends on the band dispersion. If the two bands disperse in opposite direction, the first term in the square brackets yields a negative contribution to the SW, whereas the second and third (nodal) contributions are positive. This SW relies solely on the intraband velocity and is therefore independent of the Bloch quantum geometry. Consequently, even a model without quantum geometric effects can potentially exhibit a negative SW and may therefore stabilize a PDW order. This is the scenario investigated in Ref.~\onlinecite{wang2025negativeroutepairdensity}.




\section{The Fulde-Ferrel PDW}
\label{sec:FFPDW}
\begin{figure}
    \centering
    \includegraphics[width=\columnwidth]{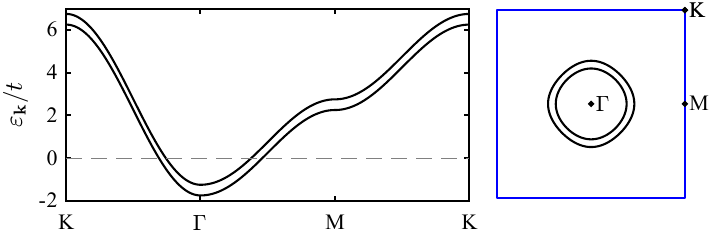}
    \caption{The left panel displays the Zeeman-split band structure of the model in Eq.~\eqref{eq:zdispersion} with parameters $(\mu,h)=(-2.5,0.25)t$, while the right panel shows the corresponding Fermi surfaces. }
    \label{fig:zeeman_band}
\end{figure}

We first study the classical model of $s$-wave pairing subject to external Zeeman splitting field. Investigating the SW of this model helps to elucidate the connection between the SW with the formation of the FF-type of PDW state. We construct the model on a square lattice with nearest-neighbor hopping $t$ and a Zeeman-splitting field $h$. The kinetic Hamiltonian $H$ can be written in the basis $\psi_{\bk}=(c_{\bk\ua},c_{\bk\da})^T$ as $H=\sum_{\bk}\psi_{\bk}^{\dagger}H_{\bk}\psi_{\bk}$. The $2\times2$ matrix $H_{\bk}$ is given by
\begin{equation}\label{eq:zdispersion}
    H_{\bk}=\varepsilon_{\bk}\mathbf{I}+h\sigma_z,
\end{equation}
with $\varepsilon_{\bk}=-2t(\cos{k_x}+\cos{k_y})-\mu$, and $\sigma_z$ acts in spin space. Fig~\ref{fig:zeeman_band} shows the band structure and Fermi surfaces under a set of parameters used for our representative calculations. Clearly, this model can also be viewed as a spinless interband $s$-wave pairing model with the two bands split by a constant energy. The BdG Hamiltonian can be written in the Nambu basis $\Psi_{\bk}=(c_{\bk\ua},c_{\bk\da},c_{\bar{\bk}\ua}^{\dagger},c_{\bar{\bk}\da}^{\dagger})^T$ as $H^{\text{BdG}}=\sum_{\bk}\Psi_{\bk}^{\dagger}H_{\bk}^{\text{BdG}}\Psi_{\bk}$, with
\begin{equation}
    H^{\text{BdG}}_{\bk}=
    \begin{pmatrix}
        H_{\bk} && \bDelta\\
        \bDelta^{\dagger} && -H_{\bar{\bk}}^*
    \end{pmatrix},
\end{equation}
where $\bDelta=\Delta i\sigma_y$, with $\Delta$ taken to be real for simplicity. The corresponding quasiparticle energy spectrum is
\begin{equation}
    \begin{aligned}
        E_{\alpha,\bk}^{\pm}&=h\pm\sqrt{\varepsilon_{\bk}^2 + \abs{\Delta}^2},\\
        E_{\beta,bk}^{\pm}&=-h\pm\sqrt{\varepsilon_{\bk}^2 + \abs{\Delta}^2}.
    \end{aligned}
\end{equation}
The model is fully gapped for $\Delta >h$, while it exhibits Bogoliubov Fermi surfaces when $\Delta \leq h$.\par

\begin{figure}
    \centering
    \includegraphics[width=\columnwidth]{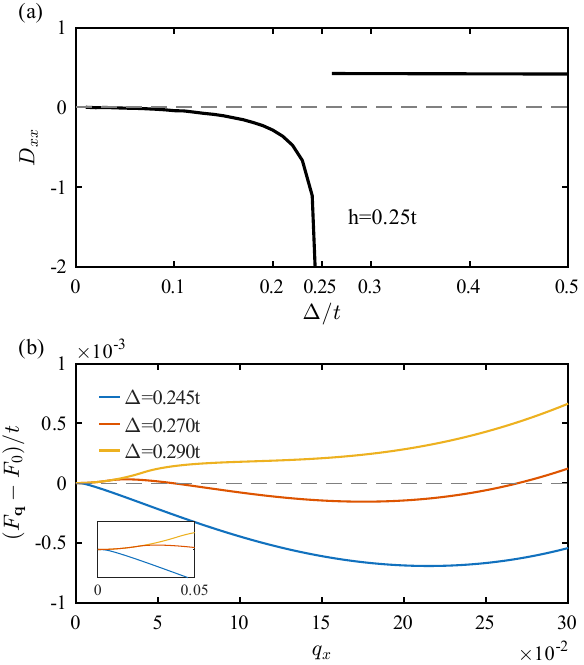}
    \caption{(a) The zero-temperature superfluid weight $D_{xx}$ of the two-band model in Fig.~\ref{fig:zeeman_band}. The curve exhibits a discontinuity at $\Delta=h=0.25t$. (b) Variation of the non-self-consistent free energy difference $F_{\bq}-F_0$ along the $q_x$ direction for three different $\Delta$. The inset magnifies the curve near $q_x=0$.  }
    \label{fig:zeeman_sw}
\end{figure}

Fig~\ref{fig:zeeman_sw} (a) presents the zero-temperature SW as a function of the pairing gap $\Delta$, evaluated according to Eq.~\eqref{eq:SW_formula}. In the current model, the SW originates entirely from the term associated with $\abs{\Delta_{12}}^2 V^{11}_{\mu}\bar{V}^{22}_{\nu}$. Since $V^{11}V^{22}\geq 0$ everywhere in momentum space, this contribution is positive in gapped area and negative at wavevectors coinciding with the Bogoliubov Fermi surface. The total SW is positive for $\Delta > h $, in which regime the Bogoliubov spectrum remains fully gapped. It changes abruptly and turns negative at the critical value of $\Delta_c = h \approx 0.25t$, below which there exists a Bogoliubov Fermi surface. Thermal excitations at finite temperatures typically further suppress the SW.

As a preliminary confirmation of the result in Fig.~\ref{fig:zeeman_sw} (a), we select three representative values of $\Delta$ spanning a range of gap size encompassing the critical pairing amplitude $\Delta_c$, and evaluate the non-self-consistent (i.e., with $\Delta$ held fixed) zero-temperature mean-field free energy for pairings with different center-of-mass momentum $\bq$, $\Delta e^{i\bq\cdot\br}$. The free energy is calculated according to
\begin{equation}
    \begin{aligned}
    F_\bq&=\frac{1}{\Omega}\sum_{\mk}\sum_{E<0}E_{\lambda,\mk,\bq}^{\pm}+\frac{1}{\Omega}\sum_{\mk}2\varepsilon_{\bk}+\frac{1}{\Omega}\sum_{\mk}\frac{\Delta^2}{V},
    \end{aligned}
\end{equation}
where $E_{\lambda,\mk,\bq}^{\pm}$ is the quasiparticle spectrum with $\lambda=\alpha,\beta$. Figure~\ref{fig:zeeman_sw} (b) presents the free energy difference between the PDW and the normal states as a function of the pairing modulation wavevector $\bq$. The three curves exhibit three distinct evolution patterns. In the regime where $\Delta > h$, the free energy displays a global minimum at $\bq=0$ and rises monotonically with $\bq$, implying a positive SW and a stable uniform superconducting pairing. In the regime with $\Delta < h$, the free energy exhibits a local maximum at $\bq=0$, consistent with the negative SW obtained in Fig~\ref{fig:zeeman_sw}~(a), suggesting that uniform pairing is unstable. In a narrow intermediate window immediately above $\Delta_c$, approximately in the range $\Delta \in (0.255t,0.275t]$ in the current calculation, the free energy curve initially increases and then drops with increasing $\bq$, until it rises again beyond a larger value of $\bq$. As a result, the free energy develops a local minimum at $\bq=0$ and a global minimum at some finite $\bq$. Hence, while the uniform pairing state exhibits positive SW, the most stable pairing is a PDW. The three representative free energy curves do reproduce the three scenarios sketched in Fig.~\ref{fig:three_modes}, thereby establishing a preliminary connection between the sign of the SW and the formation of PDW order. 

However, the above results must be treated with caution, as they were obtained from non-self-consistent calculations which do not account for the possible variation of the pairing amplitude with modulation wavevector. Furthermore, these calculations ignored the possibility of a discontinuous evolution of the superconducting order parameter with $\bq$. Thus, to obtain a more accurate phase diagram, in the following we determine the pairing gap self-consistently for each value of $\bq$. 


\begin{figure*}
    \centering
    \includegraphics[width=0.7\linewidth]{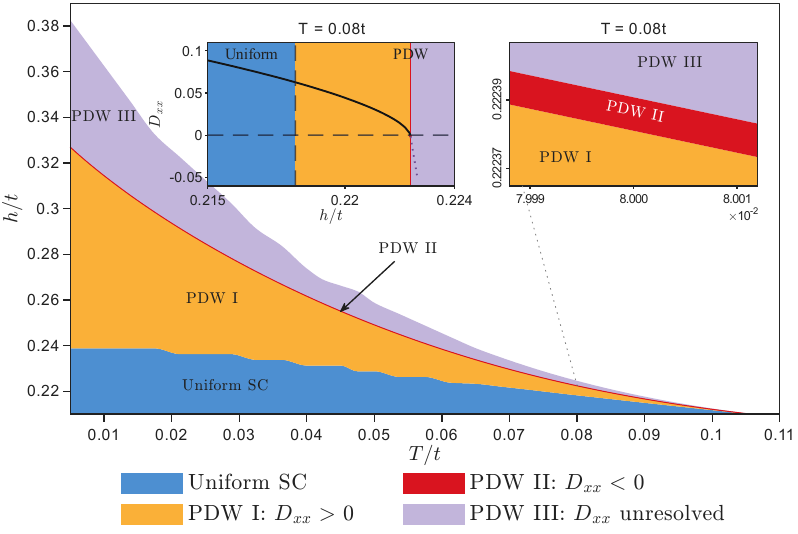}
    \caption{Self-consistent mean-field phase diagram of the two-band model in the $T$-$h$ plane, with chemical
potential $\mu=-2.5t$ and interaction strength $V=3t$. The colors encode the mean-field ground state and the corresponding superfluid weight at $\bq=0$: uniform superconductivity with $D_{xx}>0$ (blue); PDW state with positive $D_{xx}$ (orange), negative $D_{xx}$ (red), and unresolved $D_{xx}$ at $\bq=0$ (purple). The SW $D_{xx}$ is obtained from the curvature $F''_{\bq=0}$ of the self-consistent free energy. In the purple regime, self-consistent calculations cannot obtain finite superconducting gap around $\bq=0$. The left inset shows $D_{xx}$ as a function of $h$ at $T=0.08t$, with shaded backgrounds distinguishing the uniform and PDW regimes (PDW I, PDW II and PDW III). The extrapolated dotted line illustrates the trend of $D_{xx}$ in the range of larger $h$ which correspond to the purple-shaded region in the main phase diagram. The right inset zooms in on the narrow negative SW region at the same temperature.}
    \label{fig:gra_phase_diagram}
\end{figure*}

\begin{figure}
    \centering
    \includegraphics[width=\columnwidth]{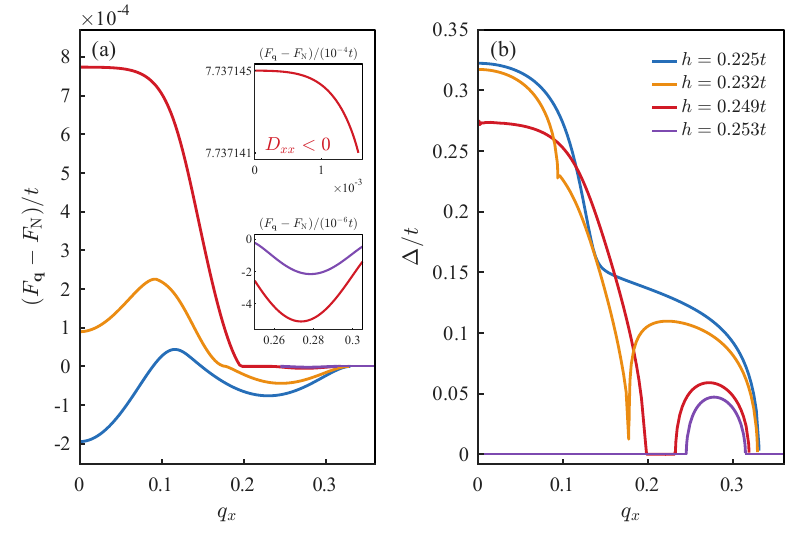}
    \caption{(a) Representative free-energy curves along the $q_x$ direction and (b) the corresponding order parameter variations, selected from Fig.~\ref{fig:gra_phase_diagram}, at temperature $T=0.05t$. The superconducting free-energy in (a) is subtracted by the normal-state free energy $F_N$. The upper (lower) inset in (a) shows the magnified view of the free energy near $q_x=0$ ($q_x=0.28$) for the red (red and purple) curve(s) in the main figure.}
    \label{fig:zeeman_scF}
\end{figure}

The $s$-wave gap equation for finite $\bq$ is given by
\begin{equation}
    \Delta_{\bq}=-\frac{V}{\Omega}\sum_{\bk}
        \left\langle c_{-\bk+\bq,\da}c_{\bk\ua}\right\rangle.
\end{equation}
For simplicity, we neglect the minor correction to the particle number due to superconducting pairing. In other words, we hold the chemical potential fixed during the self-consistent iterations. At a fixed pairing interaction strength $V=3t$, we scan through two tuning parameters, magnetic field $h$ and temperature $T$, and determine the self-consistent order parameter $\Delta_\bq$ and free energy $F_\bq$ for each $\bq$. For each set of $h$ and $T$, the SW (at $\bq=0$) is extracted using Eq.~\eqref{eq:Fq2ndderivative}. The phase diagram is shown in Fig.~\ref{fig:gra_phase_diagram}. The mean-field ground state and the sign of the SW is color-coded in different shaded regions. The results agree qualitatively with the above non-self-consistent analysis, but with some noticeable distinctions. 

Representative curves of $F_\bq$ and $\Delta_\bq$ as a function of $\bq$ are displayed in Fig.~\ref{fig:zeeman_scF}. Only the modulation wavevector range $q_x\leq 0.4$ is plotted, because the pairing order parameter vanishes at larger $q_x$ in the present model. As can be seen, with increasing magnetic field, the curvature of the free energy curve at $q=0$ turns from positive ($D>0$) to negative ($D<0$). It is important to note that the gap indeed varies drastically with $\bq$. Furthermore, for field strength that stabilizes PDW order, the self-consistent gap and the free-energy curve typically experience a discontinuous change at some $h$-dependent finite wavevectors. 

Overall, the SW is suppressed with increasing field, as can be seen in the left inset of Fig.~\ref{fig:gra_phase_diagram}. However, the PDW phase (`PDW I') onsets before the SW turns negative. This field-induced first-order phase transition is associated with the relative balance between the initial free energy minimum at $\bq=0$ and a second minimum at $\bq\sim h/v_F$ (see Fig.~\ref{fig:zeeman_scF}). Furthermore, while the trend of the SW decreasing towards negative regime is obvious in the left inset of Fig.~\ref{fig:gra_phase_diagram}, negative SW (i.e., downward free energy curve at $\bq=0$) is numerically achieved only in an extremely narrow range of field strength (region labeled `PDW II' in Fig.~\ref{fig:gra_phase_diagram}). Above this narrow range of field, in the purple-shaded regime (`PDW III') in the phase diagram, self-consistent calculations cannot obtain finite superconducting order near $\bq=0$ (see Fig.~\ref{fig:zeeman_scF}). Nonetheless, finite order parameter can be obtained in a regime of larger $\bq$ that encloses the ground-state PDW wavevector (see lower inset of Fig.~\ref{fig:zeeman_scF}). Finally, as can be seen in the left inset of Fig.~\ref{fig:gra_phase_diagram}, the rate of the field-suppression increases drastically around the same narrow field range. This signals the appearance of Bogoliubov Fermi surfaces, which would incur a large negative contribution to the SW according to Eq.~\eqref{eq:convSW}.

In summary, our above analysis has provided a more microscopic perspective into the formation of the FF-type PDW phase in $s$-wave superconductors subject to Zeeman splitting field. The PDW order in this system develops in conjunction with the field-induced progressive suppression of the SW. Notably, the suppression accelerates sharply once extensive gapless excitations, i.e., Bogoliubov Fermi surfaces, start to emerge at large field strength. 

\section{Honeycomb $f$-wave model}
\label{sec:Honeycomb}
We now proceed to the spinless honeycomb-lattice model with only nearest-neighbor hopping and nearest-neighbor $f$-wave pairing. A modified version of this model has been investigated by a sophisticated DMRG study that finds a PDW instability at around half filling~\cite{Yifan2024Pair}. In the present study, we show that this PDW may be qualitatively explained by the negative SW generated by interband Cooper pairing between oppositely dispersing bands.  

\begin{figure}
    \centering
    \includegraphics[width=\linewidth]{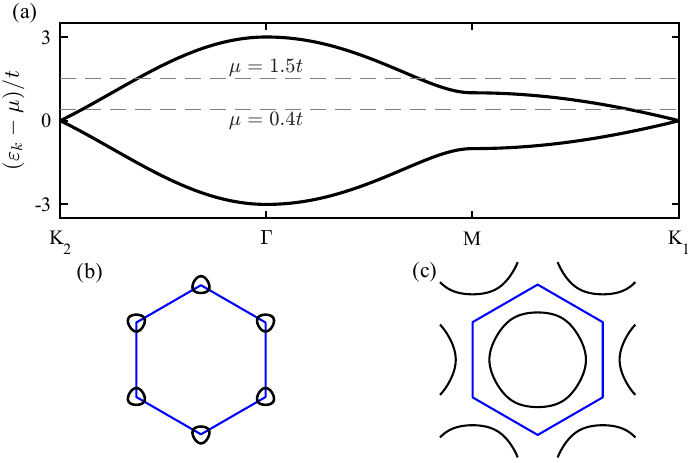}
    \caption{(a) The band structure of the honeycomb lattice model in Eq.~\eqref{eq:graphene_Hk}, with dashed lines indicating the Fermi level at chemical potential $\mu=0.4t$ and $1.5t$. (b) and (c) show the corresponding Fermi surfaces at $\mu=0.4t$ and $\mu=1.5t$, respectively, where the blue hexagon denotes the first Brillouin zone.}
    \label{fig:graphene_band}
\end{figure}

In the spinor basis $\psi_{\bk}=(c_{A\mk},c_{B\mk})^T$ where $c^\dagger_{A(B),\bk}$ and $c_{A(B),\bk}$ denote the creation and annihilation of electrons on sublattice A(B) of the honeycomb lattice, the kinetic Hamiltonian of the model follows as $H = \sum_{\bk} \psi^\dagger_{\bk} H_{\bk} \psi_{\bk}$, with
\begin{equation}\label{eq:graphene_Hk}
    H_{\bk}=
    \begin{pmatrix}
        -\mu && t\Gamma_{\bk}\\
        t\Gamma_{\bk}^* && -\mu
    \end{pmatrix}.
\end{equation}
Here $\mu$ is the chemical potential for both A and B sublattices, $t$ is the hopping between nearest sublattices, $\Gamma_{\bk}=\sum_i e^{-i\bk\cdot e_i}$ with the lattice vectors defined as $e_{1,2}=(\frac{1}{2},\pm\frac{\sqrt{3}}{2}),~e_{3}=(-1,0)$. Figure~\ref{fig:graphene_band} shows the energy dispersion and the Fermi surfaces for $\mu=0.4t$ and $\mu=1.5t$.

We introduce the following four-fermion interaction that would induce inter-sublattice pairing
\begin{equation}\label{eq:gra_int}
\begin{aligned}
    &H^{\text{int}}=-\frac{V}{\Omega}\times\\
    &\quad \sum_{\bq,\mk,\mk'}\eta_{\mk}\eta_{\mk'}^*c^{\dagger}_{A,\mk+\bq/2}c^{\dagger}_{B,\bar{\mk}+\bq/2}c_{B,\bar{\mk}'+\bq/2}c_{A,\mk'+\bq/2},
\end{aligned}
\end{equation}
where $\eta_{\mk}$ stands for the form factor of the pairing. After mean-field decoupling, the corresponding BdG Hamiltonian is written in the Nambu basis
\begin{equation}
    \Psi_{\bk,\bq}= (c_{A,\bk+\bq/2},c_{B,\bk+\bq/2},c_{A,\bar{\bk}+\bq/2}^{\dagger},c_{B,\bar{\bk}+\bq/2}^{\dagger})^T
\end{equation}
as $H^\text{BdG}_{\bq} = \sum_{\bk} \Psi^\dagger_{\bk,\bq}H^\text{BdG}_{\bk,\bq}\Psi_{\bk,\bq}$, with
\begin{equation}\label{eq:graphene_bdg}
    H^{\text{BdG}}_{\bk,\bq}=
    \begin{pmatrix}
        H_{\bk+\bq/2} && \bDelta_{\bk,\bq}^{\text{sl}}\\
        \bDelta^{\text{sl}\dagger}_{\bk,\bq} && -H_{\bar{\bk}+\bq/2}^*
    \end{pmatrix}.
\end{equation}
Here, the sublattice-basis pairing matrix follows as,
\begin{equation}
    \bDelta_{\bk,\bq}^{\text{sl}}=
    \Delta_{\bq} \begin{pmatrix}
        0 && \eta_\bk \\
        -\eta_{\bk}^* && 0
    \end{pmatrix}.
\end{equation}
The superconducting order parameter can be self-consistently determined according to the gap equation
\begin{equation}
    \Delta_{\bq}=-\frac{V}{\Omega}\sum_{\bk^\prime}\eta_{\bk^\prime}^*\left\langle c_{B,\bar{\mk}^\prime+\bq/2} c_{A,\bk^\prime+\bq/2}\right\rangle.
    \label{eq:sfdelta}
\end{equation}
In the present study, we focus on the nearest-neighbor $f$-wave pairing whose form factor reads $\eta_{\bk}=\Gamma_{\bk}/\sqrt{3}$. To check the form of the uniform pairing in the band basis, we transform the BdG Hamiltonian at $\bq=0$ using the following unitary matrix that diagonalizes the kinetic Hamiltonian $H_\bk$, 
\begin{equation}\label{eq:graphene_Uk}
    U_{\bk}=\frac{1}{\sqrt{2}}
    \begin{pmatrix}
        \frac{\Gamma_{\bk}}{\abs{\Gamma_{\bk}}} && -\frac{\Gamma_{\bk}}{\abs{\Gamma_{\bk}}}\\
        1 && 1
    \end{pmatrix}.
\end{equation}
The resultant band-basis pairing follows as
\begin{equation}
    \bDelta_{\bk}=U_{\bk}^{-1}\bDelta_{\bk,\bq=0}^{\text{sl}} U_{\bk}=\frac{\Delta}{\sqrt{3}}
    \begin{pmatrix}
        0 && \abs{\Gamma_{\bk}} \\
        -\abs{\Gamma_{\bk}} && 0
    \end{pmatrix}.
\end{equation}
It is interesting to see that only interband pairing is present. This has profound consequence on the fate of the pairing. 

Following a systematic derivation in Ref.~\onlinecite{Hu2025Quantumgeometric}, the SW of this model is explicitly derived in the Appendix~\ref{sec:app_SW}. There, it is shown that, due to a unique matching between the form of the kinetic hopping terms and the sublattice-basis pairing gap function, the quantum geometric contribution vanishes exactly (see Eqs.~\eqref{eq:app_geo} and \eqref{eq:app_honeycomb_cancellation}). Moreover, the contribution associated with the nodal excitations in the second and third terms in Eq.~\eqref{eq:sw_interband} also happens to exactly cancel another contribution related to the gradient of the pairing form factor (see Eqs.~\eqref{eq:app_conv} and \eqref{eq:app_conv_nodal}). The net zero-temperature SW thus results entirely from the first term in Eq.~\eqref{eq:sw_interband}, which is negative thanks to the exactly opposite dispersion of the two bands. We numerically verify the negative SW under two representative filling conditions, $\mu=0.4t$ and $\mu=1.5t$, corresponding, respectively, to scenarios with small Fermi surfaces near the Dirac point and a generic, large Fermi surface, as shown in Fig.~\ref{fig:graphene_band} (b) and (c). The variation of the SW as a function of the pairing amplitude, obtained from the analytical expressions in the Appendix~\ref{sec:app_SW}, are shown in Fig.~\ref{fig:gra_SW} (a) and (c). As can be seen, the SW invariably remains negative throughout the parameter space. This can also be inferred from the curvature of the free-energy near zero modulation wavevector $\bq$ (Fig.~\ref{fig:gra_SW} (b) and (d)). Furthermore, similar to the model in the previous section, the SW displays a sharp transition at some model-dependent critical strength $\Delta_c$ (see Appendix~\ref{sec:app_SW} for the explicit expression), below which a Bogoliubov Fermi surface emerges in the quasiparticle spectra.

\begin{figure}[htbp]
    \includegraphics[width=\linewidth]{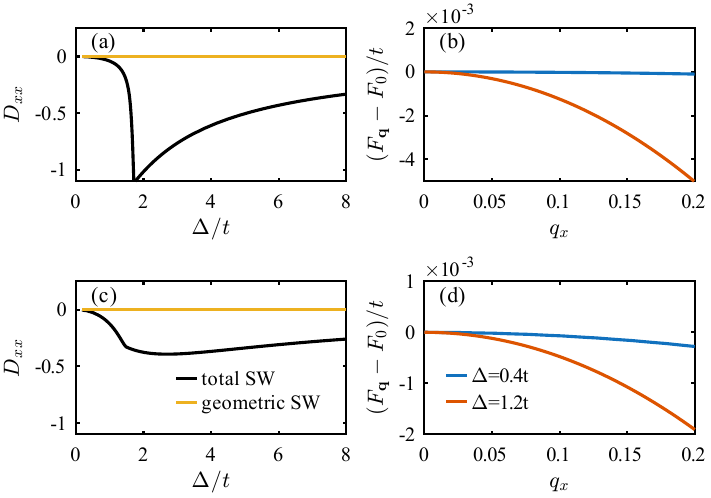}
    \caption{(a) and (c) The non-self-consistent SW as a function of pairing strength $\Delta$ in the honeycomb $f$-wave model at $\mu=0.4t$ and $1.5t$, respectively. Black and yellow curves denote, respectively, the total SW and the geometric contributions. The geometric contribution vanishes in this model. (b) and (d) Non-self-consistent free energy as a function of the Cooper pair momentum $q_x$ for $\mu=0.4t$ and $1.5t$, respectively. In both (b) and (d), the blue and red curves correspond to $\Delta=0.4t$ and $1.2t$, respectively.}
    \label{fig:gra_SW}
\end{figure}

\begin{figure}
    \centering
    \includegraphics[width=\columnwidth]{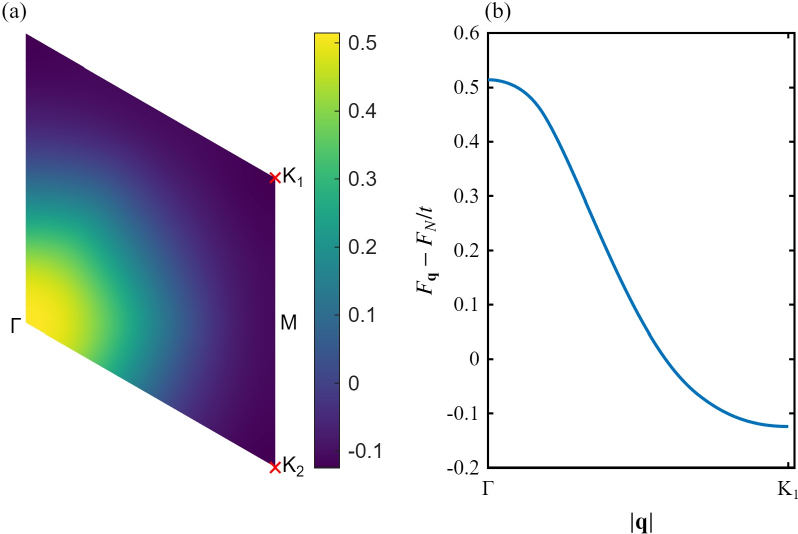}
    \caption{The self-consistent free energy $F_{\bq}$ of the model in Fig.~\ref{fig:graphene_band} with $\mu=0.4t$ and interaction strength $V=4.9t$. The normal-state energy $F_N$ is subtracted. (a) The free-energy color map in the $\bq$-space. The red cross symbols mark the position of the global free energy minima. Only one-third of the first Brillouin zone is shown, as the remaining portion is related by $C_3$ rotation symmetry. (b) Free energy profile along the $\Gamma$-$K_{1}$ direction.}
    \label{fig:gra_scf}
\end{figure}

To consolidate the above conclusion, we further perform self-consistent calculations based on Eq.~\eqref{eq:sfdelta}. For each modulation wavevector $\bq$, we self-consistently determine the pairing amplitude $\Delta_{\bq}$ and then evaluate the corresponding free energy $F_{\bq}$. The computations are carried out over the chemical potential range $\mu \in [-1.5t,1.5t]$, and representative results for $\mu=0.4t$ and $V=4.9t$ are shown in Fig.~\ref{fig:gra_scf}. Consistent with the non-self-consist results, the free energy is a local maximum at $\bq=0$, i.e.,~$D<0$, in all cases studied. Intriguingly, the global free-energy minimum always occurs at high-symmetry wavevector $\boldsymbol{K}$ and $-\boldsymbol{K}$, which is roughly consistent with Ref.~\onlinecite{Yifan2024Pair}. We caution, however, that while the self-consistent analysis may capture the correct qualitative behavior near $\bq=0$, its predictive power may fail at large $\bq$---where lattice effects become prominent. Hence, while one can conclude with confidence the tendency toward PDW order if Cooper pairing develops in the inter-sublattice $f$-wave channel, the exact modulation wavevector shall be determined by the more reliable many-body approaches. 

We also remark on the difference between our model and the model in Ref.~\onlinecite{Yifan2024Pair}. Most importantly, the two models consider different pairing interactions. In the present study, we assume only a pairing interaction in the inter-sublattice $f$-wave channel, Eq.~\eqref{eq:gra_int}, and our mean-field calculations do not consider other competing instabilities. In contrast, in Ref.~\onlinecite{Yifan2024Pair}, the interaction likely responsible for the pairing is a density-density attraction between nearest-neighbor sublattices. Their interaction could potentially promote other electronic orders, such as charge density waves, in addition to superconductivity. So far as superconductivity is concerned, their interaction may drive both $f$-wave pairing as in the present work, as well as $p$-wave pairing with $\eta_{x\bk} = (e^{i\bk\cdot \be_1}-e^{i\bk\cdot \be_2})/\sqrt{2}$ and $\eta_{y\bk}=(e^{i\bk\cdot \be_1}+e^{i\bk\cdot \be_2}-2e^{i\bk\cdot \be_3})/\sqrt{6}$. When transformed into band basis, this $p$-wave pairing has both intraband and interband Cooper pairings. Hence, it does not necessarily exhibit net negative SW, which in turn may result in a lack of robust PDW instability. The same holds for another $f$-wave channel, which involves further neighbour pairing. 

\section{Summary}
\label{sec:summary}
In this study, we have investigated the SW in systems with pure interband Cooper pairing and its implication on PDW instabilities. Our aim here is to examine this problem through the lens of the SW and to establish a qualitative link between the behavior of the SW---whether suppressed or negative---and the propensity for PDW order.

We studied two canonical models that exhibit pure interband pairing. The first is a Zeeman-split $s$-wave superconductor, which is characterized by an interband pairing between two spin-split bands. We found that, while the FF-type PDW order develops before the SW becomes negative, the SW exhibits a clear trend of increasing suppression as a function of increasing Zeeman field. In particular, it drastically turns negative when Bogoliubov Fermi surfaces emerge, signaling a clear PDW instability. The second model is a spinless honeycomb lattice with nearest-neighbor $f$-wave pairing. Owing to the peculiar form of the electronic band structure and the pairing gap function, both the quantum-geometric contribution to the SW and the conventional SW associated with nodal quasiparticle excitations are shown to vanish exactly at zero temperature. The remaining, conventional SW contribution is invariably negative due to the opposite dispersion of the two bands. This negative SW thus provides a natural explanation for the PDW instability observed in DMRG simulations of a related model~\cite{Yifan2024Pair}. 

Taken together, our results establish the SW as a useful qualitative indicator of intrinsic PDW tendencies. In particular, if the net SW is negative, uniform Cooper pairing is doomed unstable, and a PDW order shall be formed if the system were to remain superconduct. Finally, while our conclusions based on mean-field analysis may be valid in some limits of strongly correlated models, it would be of interest to investigate how quantum fluctuations and potential competing instabilities may influence the SW.

\begin{acknowledgments}
We acknowledge helpful discussions with Hao-Xin Wang and Hong Yao. This work is supported by the NSFC under Grant No.~12374042, the Guangdong Provincial Quantum Science Strategic Initiative (GDZX2601007), the Dongguan Key Laboratory of Artificial Intelligence Design for Advanced Materials, and a Shenzhen Science and Technology Program with Grant No.~KQTD20200820113010023. YFJ is supported in part by the National Key R\&D Program of China under Grants No. 2022YFA1402703, NSFC under Grant No. 12347107 and 12574160. Computing resources are provided by the SongShan Lake HPC Center (SSL-HPC) in Great Bay University.
\end{acknowledgments}

\appendix

\setcounter{figure}{0} 
\renewcommand{\thefigure}{A\arabic{figure}} 
\section{Superfluid weight formula of interband pairing}
\label{sec:app_SW}
Following Ref.~\onlinecite{Hu2025Quantumgeometric}, this appendix derives the conventional and geometric SW for models exhibiting pure interband pairing. We shall show that our honeycomb lattice model is characterized by negative net SW originated purely from conventional contribution, with vanishing geomertic SW. 


We start from a general orbital- (or sublattice-) basis BdG Hamiltonian, constructed on the basis of different orbitals or electrons on different sublattices,
\begin{equation}\label{eq:app_orbital_bdg}
H^{\mathrm{BdG}}_{\bk}=\begin{pmatrix}
H_{\bk}&\tilde{\bDelta}_{\bk}\\
\tilde{\bDelta}^{\dagger}_{\bk}&-H^*_{-\bk}
\end{pmatrix}.
\end{equation}
The SW can be derived from the standard linear response theory~\onlinecite{liangBandGeometryBerry2017}:
\begin{equation}\label{eq:app_SW_formula}
    D_{\mu \nu}=\frac{T}{\Omega} \sum_{\omega_n, \mk} \operatorname{Tr}\left\{\tilde{m}_{\mu \nu\bk}^{-1} \mathcal{G}_{\bk}+\mathcal{G}_{\bk}\tilde{V}_{\mu\bk} \mathcal{G}_{\bk}\tilde{V}_{\nu\bk}\right\},
\end{equation}
where $T$ represents temperature, $\mathcal{G}_{\bk}=(i\omega_n - \mathcal{H}_{\mk}^\text{BdG})^{-1}$ is the Gor'kov Green's function associated with the Bogoliubov-de Gennes (BdG) Hamiltonian, and $\omega_n =(2n+1)\pi T$ denotes the fermionic Matsubara frequency. The inverse mass tensor and the velocity operator are defined as follows,
\begin{align}
\tilde{m}_{\mu\nu\bk}^{-1}&=\left.\partial_{\mu}\partial_{\nu}\mathcal{H}_{\mk}^{\text{BdG}}\right|_{\Delta\to 0},\\
\tilde{V}_{\mu\bk}&=\left.\tau_z \partial_{\mu}\mathcal{H}_{\mk}^{\mathrm{BdG}}\right|_{\Delta\to 0},\label{eq:app_Vorbital}
\end{align}
where $\tau_{z}$ represents the Pauli matrices in the Nambu particle-hole space. Here and after, we use the shorthand $\partial_{\mu} \equiv \frac{\partial}{\partial k_{\mu}}$. Integrating the first term of Eq.~\eqref{eq:app_SW_formula} by parts and using $\partial_\mu\mathcal{G}_{\bk}=\mathcal{G}_{\bk}(\partial_\mu H^{\mathrm{BdG}}_{\bk})\mathcal{G}_{\bk}$ gives
\begin{widetext}
\begin{equation}\label{eq:app_response_parts}
D_{\mu\nu}=\frac{T}{\Omega}\sum_{\omega_n,\bk}\operatorname{Tr}\left[
\mathcal{G}_{\bk}\tilde{V}_{\mu\bk}\mathcal{G}_{\bk}\tilde{V}_{\nu\bk}
-\mathcal{G}_{\bk}\tau_z\tilde{V}_{\mu\bk}\mathcal{G}_{\bk}\tau_z\tilde{V}_{\nu\bk}
-\mathcal{G}_{\bk}\begin{pmatrix}
0&\partial_\mu\tilde{\bDelta}_{\bk}\\
\partial_\mu\tilde{\bDelta}^{\dagger}_{\bk}&0
\end{pmatrix}\mathcal{G}_{\bk}\tau_z\tilde{V}_{\nu\bk}\right].
\end{equation}
\end{widetext}
The last term retains the momentum dependence of the pairing matrix.

To express this response in the band basis, let $H_{\bk}\rvec{i_{\bk}}=\varepsilon_{i,\bk}\rvec{i_{\bk}}$, and define the hole-sector eigenvector $\rvec{\bar i_{\bk}}=\rvec{i_{-\bk}}^*$. The corresponding Nambu basis states are $\rvec{+,i_{\bk}}=(1,0)^T\otimes\rvec{i_{\bk}}$ and $\rvec{-,\bar i_{\bk}}=(0,1)^T\otimes\rvec{\bar i_{\bk}}$. In this basis, the velocity operator is $\operatorname{diag}(V_{\mu\bk},\bar V_{\mu\bk})$, with
\begin{align}
V^{ij}_{\mu\bk}&=\delta_{ij}\partial_\mu\varepsilon_{i,\bk}
+(\varepsilon_{i,\bk}-\varepsilon_{j,\bk})\vecDot{\partial_\mu i_{\bk}}{j_{\bk}},\\
\bar V^{ij}_{\mu\bk}&=-(V^{ij}_{\mu,-\bk})^*.
\end{align}
The off-diagonal velocity matrix elements with $i\neq j$ are closely tied to the interband Berry connection $i\vecDot{\partial_\mu i_{\bk}}{j_{\bk}}$. These interband velocities describe a quantum coherent charge transport between the bands, and they underlie the quantum geometric character of the Bloch states. In this regard, any contribution to the SW that involves interband velocity is attributed to quantum geometric effects. 

For a time-reversal-symmetric system, $\bar V^{ij}_{\mu\bk}=V^{ij}_{\mu\bk}$. The pairing matrix elements in band-basis are $\Delta_{ij,\bk}=\lvec{i_{\bk}}\tilde{\bDelta}_{\bk}\rvec{\bar j_{\bk}}$. And its derivatives $\partial_\mu\tilde{\bDelta}_{\bk}$ transforming as:
\begin{widetext}
\begin{equation}\label{eq:app_gap_derivative}
\partial_\mu\tilde{\bDelta}_{\bk}
=\sum_{i,j}\Bigl[
\Delta_{ij,\bk}\rvec{\partial_\mu i_{\bk}}\lvec{\bar j_{\bk}}
+\Delta_{ij,\bk}\rvec{i_{\bk}}\lvec{\partial_\mu\bar j_{\bk}}
+(\partial_\mu\Delta_{ij,\bk})\rvec{i_{\bk}}\lvec{\bar j_{\bk}}\Bigr].
\end{equation}
\end{widetext}
Let $\rvec{\psi_{a,\bk}}$ be the wavefunction of the $a$-th Bogoliubov quasiparticle, with energy $E_{a,\bk}$, satisfying $H^{\mathrm{BdG}}_{\bk}\rvec{\psi_{a,\bk}}=E_{a,\bk}\rvec{\psi_{a,\bk}}$. Its expansion in the band-basis is
\begin{equation}
    \rvec{\psi_{a,\bk}}=\sum_i\left(\phi^+_{ai,\bk}\rvec{+,i_{\bk}}+\phi^-_{ai,\bk}\rvec{-,\bar i_{\bk}}\right).\label{eq:app_quasiparticle_expansion}
\end{equation}
Here $a$ labels BdG quasiparticle bands, whereas $i$ labels normal-state bands; $\phi^+_{ai,\bk}$ and $\phi^-_{ai,\bk}$ are the expanded coefficients. The Green's function can also written in spectral representation as
\begin{equation}
\mathcal{G}_{\bk}
=\sum_a\frac{\rvec{\psi_{a,\bk}}\lvec{\psi_{a,\bk}}}{\mathrm{i}\omega_n-E_{a,\bk}}.\label{eq:app_spectral_green}
\end{equation}
Substituting the band-basis quantities into Eq.~\eqref{eq:app_response_parts} and performing the Matsubara sum produces the total SW~\cite{Hu2025Quantumgeometric}
\begin{equation}\label{eq:app_total_SW}
\begin{aligned}
D_{\mu\nu}=&\frac{2}{\Omega}\sum_{\bk,ijlm}C^{\ti{++--}}_{ijlm,\bk}
\left(V^{ij}_{\mu\bk}\bar V^{lm}_{\nu\bk}
+V^{ij}_{\nu\bk}\bar V^{lm}_{\mu\bk}\right)\\
&-\frac{2}{\Omega}\operatorname{Re}\sum_{\bk,ijlm\sigma}
C^{\ti{+-\sigma\sigma}}_{ijlm,\bk}
\left(\delta_{\sigma,+}V^{lm}_{\nu\bk}-\delta_{\sigma,-}\bar V^{lm}_{\nu\bk}\right)\\
&\times\left[\sum_p\left(
\Delta_{ip,\bk}\vecDot{\partial_\mu\bar p_{\bk}}{\bar j_{\bk}}
+\vecDot{i_{\bk}}{\partial_\mu p_{\bk}}\Delta_{pj,\bk}\right)
+\partial_\mu\Delta_{ij,\bk}\right],
\end{aligned}
\end{equation}
where the coefficients are
\begin{equation}\label{eq:app_C}
C^{\ti{\sigma_1\sigma_2\sigma_3\sigma_4}}_{ijlm,\bk}
=\sum_{a,b}\frac{f(E_{a,\bk})-f(E_{b,\bk})}{E_{a,\bk}-E_{b,\bk}}
\phi^{\sigma_1*}_{ai,\bk}\phi^{\sigma_2}_{bj,\bk}
\phi^{\sigma_3*}_{bl,\bk}\phi^{\sigma_4}_{am,\bk}.
\end{equation}
Here, $f(E)$ is the Fermi function, $\sigma=\pm$, and $\delta_{\sigma,\pm}$ are Kronecker deltas. Notice that when $E_{a,\bk}=E_{b,\bk}$, the fraction in the coefficient must be understood as the derivative $f'(E)$.

Equation~\eqref{eq:app_total_SW} applies to general pairing matrices; its terms can be separated into conventional and geometric contributions according to whether they involve interband velocities.

We now specialize to two bands with only interband pairing, i.e. $\Delta_{11,\bk}=\Delta_{22,\bk}=0$. In the Nambu basis $(d_{1\bk},d_{2\bk},d_{1,-\bk}^{\dagger},d_{2,-\bk}^{\dagger})^T$, the BdG matrix becomes
\begin{equation}\label{eq:app_interband_bdg}
\mathcal{H}^{\mathrm{BdG}}_{\bk}=\begin{pmatrix}
\varepsilon_{1,\bk}&0&0&\Delta_{12,\bk}\\
0&\varepsilon_{2,\bk}&\Delta_{21,\bk}&0\\
0&\Delta_{21,\bk}^*&-\varepsilon_{1,\bk}&0\\
\Delta_{12,\bk}^*&0&0&-\varepsilon_{2,\bk}
\end{pmatrix}.
\end{equation}
It separates into the $(1,\bar2)$ and $(2,\bar1)$ blocks, labeled by $\alpha$ and $\beta$. For spin-singlet pairing, one has $\Delta_{21,\bk}=-\Delta_{12,\bk}$. For later convenience, we define the following quantities:
\begin{equation}\label{eq:app_abbreviations}
\begin{aligned}
\xi_{\bk}&=\frac{\varepsilon_{1,\bk}+\varepsilon_{2,\bk}}{2},\\
\Delta_{\alpha,\bk}&=\Delta_{12,\bk},
&\Delta_{\beta,\bk}&=\Delta_{21,\bk},\\
g_{\lambda,\bk}&=\sqrt{\xi_{\bk}^{2}+|\Delta_{\lambda,\bk}|^{2}},
&\lambda&=\alpha,\beta.
\end{aligned}
\end{equation}
The quasiparticle energies are
\begin{equation}\label{eq:app_spectra}
\begin{aligned}
E^{\pm}_{\alpha,\bk}&=\frac{\varepsilon_{1,\bk}-\varepsilon_{2,\bk}}{2}\pm g_{\alpha,\bk},\\
E^{\pm}_{\beta,\bk}&=-\frac{\varepsilon_{1,\bk}-\varepsilon_{2,\bk}}{2}\pm g_{\beta,\bk}.
\end{aligned}
\end{equation}
The corresponding quasiparticle wavefunctions are explicitly
\begin{equation}\label{eq:app_interband_wavefunctions}
\begin{aligned}
\rvec{\psi^+_{\alpha,\bk}}&=u_{\alpha,\bk}\rvec{+,1_{\bk}}+v_{\alpha,\bk}\rvec{-,\bar2_{\bk}},\\
\rvec{\psi^-_{\alpha,\bk}}&=-v_{\alpha,\bk}^*\rvec{+,1_{\bk}}+u_{\alpha,\bk}^*\rvec{-,\bar2_{\bk}},\\
\rvec{\psi^+_{\beta,\bk}}&=u_{\beta,\bk}\rvec{+,2_{\bk}}+v_{\beta,\bk}\rvec{-,\bar1_{\bk}},\\
\rvec{\psi^-_{\beta,\bk}}&=-v_{\beta,\bk}^*\rvec{+,2_{\bk}}+u_{\beta,\bk}^*\rvec{-,\bar1_{\bk}}.
\end{aligned}
\end{equation}
Where the coherence factors obey $|u_{\lambda,\bk}|^2+|v_{\lambda,\bk}|^2=1$ and
\begin{equation}\label{eq:app_coherence}
\begin{aligned}
|u_{\lambda,\bk}|^2-|v_{\lambda,\bk}|^2&=\frac{\xi_{\bk}}{g_{\lambda,\bk}},\\
u_{\lambda,\bk}v_{\lambda,\bk}^*&=\frac{\Delta_{\lambda,\bk}}{2g_{\lambda,\bk}}.
\end{aligned}
\end{equation}
These wavefunctions determine the amplitudes $\phi^{\pm}_{ai,\bk}$ in Eq.~\eqref{eq:app_C}. Next, in the interband-only case, we derive the conventional and geometric SW contributions separately.

\subsection{Conventional contribution}
\label{subsec:conv}
From Eq.~\eqref{eq:app_total_SW}, retaining the terms only containing the intraband velocity, we obtain:
\begin{equation}\label{eq:app_conv}
\begin{aligned}
D^{\mathrm{conv}}_{\mu\nu}=&\frac{2}{\Omega}\sum_{\bk}
\Bigl\{C^{\ti{++--}}_{1122,\bk}\bigl[V^{11}_{\mu\bk}\bar V^{22}_{\nu\bk}+(\mu\leftrightarrow\nu)\bigr]\\
&+C^{\ti{++--}}_{2211,\bk}\bigl[V^{22}_{\mu\bk}\bar V^{11}_{\nu\bk}+(\mu\leftrightarrow\nu)\bigr]\Bigr\}\\
&-\frac{2}{\Omega}\operatorname{Re}\sum_{\bk,i\ne j}
\Bigl[C^{\ti{+-++}}_{ijii,\bk}V^{ii}_{\nu\bk}-C^{\ti{+---}}_{ijjj,\bk}\bar V^{jj}_{\nu\bk}\Bigr]\partial_\mu\Delta_{ij,\bk}\,,
\end{aligned}
\end{equation}
where the coefficients in the first summation are
\begin{equation}\label{eq:app_conv_coefficient}
\begin{aligned}
C^{\ti{++--}}_{1122,\bk}={}&\frac{|\Delta_{12,\bk}|^2}{4g_{\alpha,\bk}^2}
\Biggl[-\frac{f(E^+_{\alpha,\bk})-f(E^-_{\alpha,\bk})}{g_{\alpha,\bk}}\\
&\qquad+f'(E^+_{\alpha,\bk})+f'(E^-_{\alpha,\bk})\Biggr],
\end{aligned}
\end{equation}
with $C^{\ti{++--}}_{2211,\bk}$ obtained by $(1,2,\alpha)\leftrightarrow(2,1,\beta)$. The second summation depends on the $\bk$-dependence of pairing, with the corresponding coefficient given by:
\begin{equation}\label{eq:app_conv_form}
\begin{aligned}
C^{\ti{+-++}}_{1211,\bk}={}&\frac{\Delta_{12,\bk}^*}{2g_{\alpha,\bk}}
\Biggl[|u_{\alpha,\bk}|^2f'(E^+_{\alpha,\bk})
-|v_{\alpha,\bk}|^2f'(E^-_{\alpha,\bk})\\
&\quad-\frac{\xi_{\bk}}{2g_{\alpha,\bk}^2}
\bigl[f(E^+_{\alpha,\bk})-f(E^-_{\alpha,\bk})\bigr]\Biggr],\\
C^{\ti{+---}}_{1222,\bk}={}&\frac{\Delta_{12,\bk}^*}{2g_{\alpha,\bk}}
\Biggl[|v_{\alpha,\bk}|^2f'(E^+_{\alpha,\bk})
-|u_{\alpha,\bk}|^2f'(E^-_{\alpha,\bk})\\
&\quad+\frac{\xi_{\bk}}{2g_{\alpha,\bk}^2}
\bigl[f(E^+_{\alpha,\bk})-f(E^-_{\alpha,\bk})\bigr]\Biggr].
\end{aligned}
\end{equation}

If $\Delta_{ij,\bk}=\Delta_{ij}$ is independent of $\bk$, the second summation vanishes, and the first summation takes the form in Eq.~\eqref{eq:sw_interband} of the main text. Especially, for the fully-gapped and zero-temperature case, the coefficient Eq.~\eqref{eq:app_conv_coefficient} is positive since $f(E^+_{\alpha,\bk})\leq f(E^-_{\alpha,\bk})$. thus the first summation of Eq.~\eqref{eq:app_conv} is negative when the two bands disperse oppositely, $V^{11}_{\mu\bk}V^{22}_{\mu\bk}<0$. The $f'(E)$ terms reduce to $-\delta(E)$ at zero temperature and contribute only at nodes or Bogoliubov Fermi surfaces. Their contribution to the first
summation is opposite in sign to the non-nodal contribution.

For a pairing function that has explicit dependence on $\bk$, the second summation must also be retained. The coefficients in Eq.~\ref{eq:app_conv_form} also contain nodal and non-nodal contributions, making the sign of total conventional SW more complicated. For the honeycomb model considered below, however, the second summation happens to contain only nodal contributions. And at zero temperature, the nodal terms from the two summations exactly cancel.

\subsection{Geometric contribution}
The terms involving interband velocities give
\begin{widetext}
\begin{equation}\label{eq:app_geo}
\begin{aligned}
D^{\mathrm{geo}}_{\mu\nu}={}&\frac{2}{\Omega}\sum_{\bk}\Bigl\{
C^{\ti{++--}}_{1212,\bk}\bigl[V^{12}_{\mu\bk}\bar V^{12}_{\nu\bk}+(\mu\leftrightarrow\nu)\bigr]
+(C^{\ti{++--}}_{1212,\bk})^*\bigl[V^{21}_{\mu\bk}\bar V^{21}_{\nu\bk}+(\mu\leftrightarrow\nu)\bigr]\Bigr\}\\
&-\frac{2}{\Omega}\operatorname{Re}\sum_{\bk,i\ne j}
\Bigl[C^{\ti{+-++}}_{iiji,\bk}V^{ji}_{\nu\bk}-C^{\ti{+---}}_{iiij,\bk}\bar V^{ij}_{\nu\bk}\Bigr]
\Bigl[\Delta_{ij,\bk}\vecDot{\partial_\mu\bar j_{\bk}}{\bar i_{\bk}}
-\Delta_{ji,\bk}\vecDot{\partial_\mu i_{\bk}}{j_{\bk}}\Bigr],
\end{aligned}
\end{equation}
\end{widetext}
where the coefficient in the first line is
\begin{align}
C^{\ti{++--}}_{1212,\bk}
&=\frac{\Delta_{\alpha,\bk}^*\Delta_{\beta,\bk}}{4g_{\alpha,\bk}g_{\beta,\bk}}
\sum_{s,s'=\pm1}ss'\,
\frac{f(E^s_{\alpha,\bk})-f(E^{s'}_{\beta,\bk})}{E^s_{\alpha,\bk}-E^{s'}_{\beta,\bk}}.\label{eq:app_geo_coe1}
\end{align}
The coefficients in the second line comprise four combinations, $C_{1121,\bk}^{\ti{+-++}}$, $C_{1112,\bk}^{\ti{+---}}$, $C_{2212,\bk}^{\ti{+-++}}$, and $C_{2221,\bk}^{\ti{+---}}$. Here we list the explicit expressions of the first two
\begin{equation}
\begin{aligned}
    &C^{\ti{+-++}}_{1121,\bk}=\\
&\qquad\frac{\Delta_{\beta,\bk}^*}{2g_{\beta,\bk}}\sum_{s,s'=\pm1}
s'\,\frac{1+s\xi_{\bk}/g_{\alpha,\bk}}{2}
\frac{f(E^s_{\alpha,\bk})-f(E^{s'}_{\beta,\bk})}{E^s_{\alpha,\bk}-E^{s'}_{\beta,\bk}}\\
&C^{\ti{+---}}_{1112,\bk}=\\
&\qquad\frac{\Delta_{\alpha,\bk}^*}{2g_{\alpha,\bk}}\sum_{s,s'=\pm1}
s\,\frac{1-s'\xi_{\bk}/g_{\beta,\bk}}{2}
\frac{f(E^s_{\alpha,\bk})-f(E^{s'}_{\beta,\bk})}{E^s_{\alpha,\bk}-E^{s'}_{\beta,\bk}},
\end{aligned}
\end{equation}
while the remaining two are obtained by exchanging $\alpha$ and $\beta$.

\subsection{Honeycomb lattice model}
\label{app:honeycomb_SW}

\begin{figure}[htbp]
    \centering
    \includegraphics[width=\columnwidth]{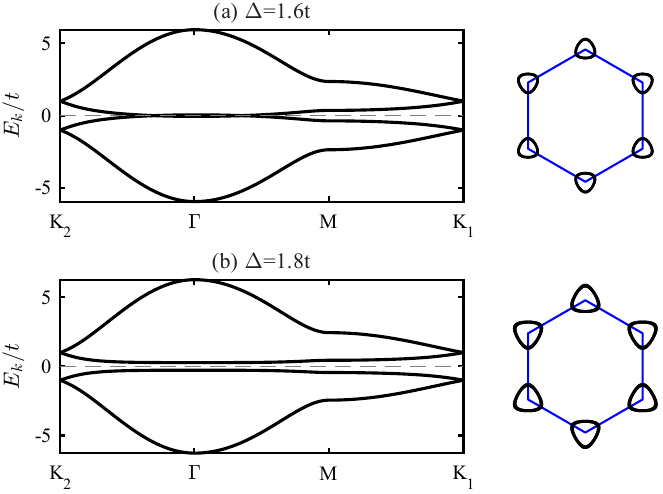}
    \caption{Quasiparticle dispersion and Fermi surface of the BdG Hamiltonian~\eqref{eq:graphene_bdg} at $\mu=0.4t$ for different pairing strength $\Delta$. A Bogoliubov Fermi surface exists in panel (a), while the gap opens in panel (b).}
    \label{fig:gra_bdgband}
\end{figure}

We now specialize to the nearest-neighbor honeycomb model Eq.~\eqref{eq:graphene_Hk} with $f$-wave pairing at zero temperature. The band dispersions and other quantities satisfy
\begin{equation}\label{eq:app_honeycomb_parameters}
\begin{aligned}
\varepsilon_{1,2,\bk}&=-\mu\pm t\abs{\Gamma_{\bk}},\\
\Delta_{12,\bk}&=-\Delta_{21,\bk}=\Delta\abs{\Gamma_{\bk}}/\sqrt{3},\\
g_{\alpha,\bk}&=g_{\beta,\bk}=\sqrt{\mu^2+\frac{1}{3}|\Delta|^2\abs{\Gamma_{\bk}}^2}.
\end{aligned}
\end{equation}
Notice that the BdG dispersion $E^-_{\alpha,\bk}=t\abs{\Gamma_{\bk}}-g_{\alpha,\bk}$, and $E^+_{\beta,\bk}=-E^-_{\alpha,\bk}$ may vanish at certain $\bk$ as governed by the condition
\begin{equation}\label{eq:app_honeycomb_nodes}
(t^2-\frac{1}{3}|\Delta|^2)\abs{\Gamma_{\bk}}^2=\mu^2.
\end{equation}
Since $\abs{\Gamma_{\bk}}\leq 3$, the critical value of $\Delta$ can be solved as
\begin{equation}\label{eq:app_honeycomb_gap_boundary}
\Delta_c=\sqrt{3t^2-\frac{\mu^2}{3}}.
\end{equation}
For $|\Delta|<\Delta_c$, there is a Bogoliubov Fermi surface, as shown in Fig.~\ref{fig:gra_bdgband}.

Using the diagonalizing matrix in Eq.~\eqref{eq:graphene_Uk}, the band-basis velocity operator, $V_{x\bk}=U_{\bk}^{\dagger}(\partial_x H_{\bk})U_{\bk}$, can be directly calculated as
\begin{widetext}

\begin{equation}\label{eq:app_honeycomb_velocity}
\begin{aligned}
V_{x\bk}&=\begin{pmatrix}
V^{11}_{x\bk}&V^{12}_{x\bk}\\[2pt]
V^{21}_{x\bk}&V^{22}_{x\bk}
\end{pmatrix}=t\begin{pmatrix}
\partial_x\abs{\Gamma_{\bk}}
&\mathrm{i}\,\dfrac{\operatorname{Im}(\Gamma_{\bk}^*\partial_x\Gamma_{\bk})}{\abs{\Gamma_{\bk}}}\\
-\mathrm{i}\,\dfrac{\operatorname{Im}(\Gamma_{\bk}^*\partial_x\Gamma_{\bk})}{\abs{\Gamma_{\bk}}}
&-\partial_x\abs{\Gamma_{\bk}}
\end{pmatrix}.
\end{aligned}
\end{equation}
Thus the velocity matrix elements satisfy
\begin{equation}\label{eq:app_honeycomb_relations}
V^{11}_{x\bk}=-V^{22}_{x\bk},\qquad
V^{12}_{x\bk}=-V^{21}_{x\bk},
\end{equation}
and $\bar V^{ij}_{x\bk}=V^{ij}_{x\bk}$ by TRS.

With these relations, one can obtain explicit expressions for the conventional and geometric contributions of $D_{xx}$. The $C$-coefficients in Eq.~\eqref{eq:app_conv} satisfy $C_{1122,\bk}^{\ti{++--}}=C_{2211,\bk}^{\ti{++--}}$ and 
\begin{equation}\label{eq:app_honeycomb_form_coefficients}
C^{\ti{+-++}}_{1211,\bk}+C^{\ti{+---}}_{1222,\bk}
=\frac{\Delta_{12,\bk}^*}{2g_{\alpha,\bk}}
\left[f'(E^+_{\alpha,\bk})-f'(E^-_{\alpha,\bk})\right],
\end{equation}
where the terms without $f'$ cancel. At zero temperature, only $E_{\alpha,\bk}^-$ can reach zero, collecting the $f'(E_{\alpha,\bk}^-)=-\delta(E_{\alpha,\bk}^-)$ terms in Eq.~\eqref{eq:app_conv} we get, for the conventional SW contribution associated with gap nodes,

    \begin{equation}\label{eq:app_conv_nodal}
    \begin{aligned}
    D_{xx}^{\text{conv-nodal}}&=\frac{2}{\Omega}\sum_{\bk}\frac{\abs{\Delta_{12,\bk}^2}}{g_{\alpha,\bk}^2}\delta(E_{\alpha,\bk}^-)(V^{11}_{x\bk})^2-\frac{2}{\Omega}\text{Re}\sum_{\bk} \frac{\Delta_{12,\bk}^*}{g_{\alpha,\bk}}\delta(E_{\alpha,\bk}^-)\partial_x \Delta_{12,\bk}V^{11}_{x\bk}\\
    &=\frac{2}{\Omega}\sum_{\bk}\left(\frac{t\abs{\Gamma_{\bk}}}{g_{\alpha,\bk}}-1 \right)\frac{\Delta^2t\abs{\Gamma_{\bk}}}{3g_{\alpha,\bk}}(\partial_x \abs{\Gamma_{\bk}})^2\delta(E_{\alpha,\bk}^-)\\
    &=0.
    \end{aligned}
\end{equation}
Here, the $\delta$-function contributes only when $E_{\alpha,\bk}^-=t\abs{\Gamma_{\bk}}-g_{\alpha,\bk}=0$, where the multiplying expression in the front, $t\abs{\Gamma_{\bk}}/g_{\alpha,\bk}-1$, also vanishes, so the nodal contribution vanishes. This nodal contribution remains particularly small for a similar mechanism even at elevated temperatures.

Below, we turn to the geometric SW. From Eq.~\eqref{eq:app_honeycomb_parameters}, we find that the coefficients in Eq.~\eqref{eq:app_geo} satisfy
\begin{equation}\label{eq:app_honeycomb_identity}
\begin{aligned}
C^{\ti{+-++}}_{1121,\bk}+C^{\ti{+---}}_{1112,\bk}
+&C^{\ti{+-++}}_{2212,\bk}+C^{\ti{+---}}_{2221,\bk}\\
&=\frac{\Delta_{12,\bk}}{2g_{\alpha,\bk}}\sum_{s,s'=\pm1}(s-s')\frac{f(E^s_{\alpha,\bk})-f(E^{s'}_{\beta,\bk})}{E^s_{\alpha,\bk}-E^{s'}_{\beta,\bk}}\\
&=\frac{\Delta_{12,\bk}}{2g_{\alpha,\bk}^2}\sum_{s,s'=\pm1}\left[ 2t\abs{\Gamma_{\bk}}- (E_{\alpha,\bk}^s-E_{\beta,\bk}^{s'}) \right] ss'\frac{f(E^s_{\alpha,\bk})-f(E^{s'}_{\beta,\bk})}{E^s_{\alpha,\bk}-E^{s'}_{\beta,\bk}}\\
&=\frac{t\abs{\Gamma_{\bk}}\Delta_{12,\bk}}{g_{\alpha,\bk}^2}\sum_{s,s'=\pm1}ss'\frac{f(E^s_{\alpha,\bk})-f(E^{s'}_{\beta,\bk})}{E^s_{\alpha,\bk}-E^{s'}_{\beta,\bk}}\\
&=-4t\frac{\abs{\Gamma_{\bk}}}{\Delta_{12,\bk}} C^{\ti{++--}}_{1212,\bk}.
\end{aligned}
\end{equation}
Here we have used $E_\alpha^s-E_\beta^{s'}
=2t|\Gamma|+(s-s')g$ and $ss'(s-s')=s'-s$.
Finally, we reach the crucial conclusion that the geometric contribution Eq.~\eqref{eq:app_geo} vanishes at each momentum:
\begin{equation}\label{eq:app_honeycomb_cancellation}
D^{\mathrm{geo}}_{xx}
=\frac{1}{\Omega}\sum_{\bk}
\left\{8C^{\ti{++--}}_{1212,\bk}
+\frac{2\Delta_{12,\bk}}{t\abs{\Gamma_{\bk}}}
\Bigl[C^{\ti{+-++}}_{1121,\bk}+C^{\ti{+---}}_{1112,\bk}
+C^{\ti{+-++}}_{2212,\bk}+C^{\ti{+---}}_{2221,\bk}\Bigr]\right\}
(V^{12}_{x\bk})^2=0,
\end{equation}
where we have used $V^{12}_{x\bk}=-V^{21}_{x\bk}$. This cancellation holds at finite temperatures as well.

With the geometric and nodal-conventional contributions both vanishing, the remaining zero-temperature SW is
\begin{equation}\label{eq:app_honeycomb_remaining_SW}
D_{xx}=D^{\mathrm{conv}}_{xx}
=\frac{2}{\Omega}\sum_{\bk}
\frac{1}{g_{\alpha,\bk}^3}
\left[f(E_{\alpha,\bk}^-)-f(E_{\alpha,\bk}^+)\right] V^{11}_{x\bk}V^{22}_{x\bk}\Delta_{12,\bk}^*\Delta_{12,\bk}\,\leq0,
\end{equation}
\end{widetext}
consistent with the numerical results in Fig.~\ref{fig:gra_SW}. At finite temperature, this term remains negative, and since the nodal contribution is rather small, the total SW is expected to remain negative.

\bibliography{huyijian}

\end{document}